\documentclass[aps,notitlepage,amsmath,amssymb]{revtex4-1}

\begin{document}

\title{Canonical ensemble in non-extensive statistical mechanics when $q>1$}

\author{Julius Ruseckas}
\email{julius.ruseckas@tfai.vu.lt}
\homepage{http://www.itpa.lt/~ruseckas}
\affiliation{Institute of Theoretical Physics and Astronomy, Vilnius University,
A.~Go\v{s}tauto 12, LT-01108 Vilnius, Lithuania}

\begin{abstract}
The non-extensive statistical mechanics has been used to describe
a variety of complex systems. The maximization of entropy, often used
to introduce the non-extensive statistical mechanics, is a formal
procedure and does not easily leads to physical insight. In this article
we investigate the canonical ensemble in the non-extensive statistical
mechanics by considering a small system interacting with a large reservoir
via short-range forces and assuming equal probabilities for all available
microstates. We concentrate on the situation when the reservoir is
characterized by generalized entropy with non-extensivity parameter
$q>1$. We also investigate the problem of divergence in the non-extensive
statistical mechanics occurring when $q>1$ and show that there is
a limit on the growth of the number of microstates of the system that
is given by the same expression for all values of $q$.
\end{abstract}

\maketitle

\section{Introduction}

The standard, Boltzmann-Gibbs statistical mechanics has been successfully
applied to describe a huge variety of systems. The cornerstone of
the standard statistical mechanics is the functional form of the entropy
\begin{equation}
S_{\mathrm{BG}}=-k_{\mathrm{B}}\sum_{\mu}p(\mu)\ln p(\mu)\,,\label{eq:BG}
\end{equation}
where $p(\mu)$ is the probability of finding the system in the state
characterized by the parameters $\mu$. However, there are systems
exhibiting long-range interactions, long-range memory, and anomalous
diffusion, that possess anomalous properties in view of traditional
Boltzmann-Gibbs statistical mechanics. To understand such systems
a generalization of statistical mechanics has been proposed by Tsallis
\cite{Tsallis2009-1}. The non-extensive statistical mechanics has
been used to describe phenomena in many physical systems: dusty plasmas
\cite{Liu2008}, trapped ions \cite{DeVoe2009}, spin-glasses \cite{Pickup2009},
anomalous diffusion \cite{Huang2010,Prehl2012}, high-energy physics
\cite{Adare2011}, Langevin dynamics with fluctuating temperature
\cite{Budini2012,Du2012}, cold atoms in optical lattices \cite{Lutz2013},
turbulent flows \cite{Beck2013}. This generalized framework has found
applications also in chemistry, biology, geology, and economics \cite{Gell-Mann2004,Abe2006,Picoli2009,Vallianatos2013}.
Instead of Eq.~(\ref{eq:BG}) the non-extensive statistical mechanics
is based on the generalized functional form of the entropy \cite{Tsallis2009-1}
\begin{equation}
S_{q}=k_{\mathrm{B}}\frac{1-\sum_{\mu}p(\mu)^{q}}{q-1}\,.\label{eq:q-entr}
\end{equation}
Here the parameter $q$ describes the non-extensiveness of the system.
The Boltzmann-Gibbs entropy can be obtained from Eq.~(\ref{eq:q-entr})
in the limit $q\rightarrow1$ \cite{Tsallis2009-1,Tsallis2009-2}.
More generalized entropies and distribution functions are introduced
in Refs.~\cite{Hanel2011-1,Hanel2011-2}.

It is convenient to write the equations of non-extensive statistical
mechanics using the $q$-logarithm 
\begin{equation}
\ln_{q}x=\frac{x^{1-q}-1}{1-q}\label{eq:q-log1}
\end{equation}
and its inverse, the $q$-exponential \cite{Tsallis2009-1}
\begin{equation}
\exp_{q}(x)\equiv[1+(1-q)x]_{+}^{\frac{1}{1-q}}\,.\label{eq:q-exp1}
\end{equation}
Here $[x]_{+}=x$ if $x>0$, and $[x]_{+}=0$ otherwise. For example,
using the $q$-logarithm one can write Eq.~(\ref{eq:q-entr}) in
a form similar to the Boltzmann-Gibbs entropy (\ref{eq:BG}) \cite{Tsallis2009-1}:
\begin{equation}
S_{q}=k_{\mathrm{B}}\sum_{\mu}p(\mu)\ln_{q}\frac{1}{p(\mu)}\,.
\end{equation}
The exponential Boltzmann factor in the non-extensive statistical
mechanics is replaced by a $q$-exponential. In the limit $q\rightarrow1$
the $q$-logarithm becomes an ordinary logarithm and the $q$-exponential
function becomes the ordinary exponential $e^{x}$.

In the non-extensive statistical mechanics the canonical ensemble
is often described in a formal way, starting from the maximization
of the generalized entropy (\ref{eq:q-entr}) \cite{Tsallis2009-1}.
The physical content enters as a form of constraints in the maximization
procedure. In Ref.~\cite{Ruseckas2016} the canonical ensemble in
the non-extensive statistical mechanics has been considered starting
from a physical situation of a small system interacting with a large
reservoir via short-range forces. Assuming that the $q$-heat capacity
of the reservoir instead of the ordinary heat capacity is large, the
equations of the non-extensive statistical mechanics have been obtained.
However, in Ref.~\cite{Ruseckas2016} only the case of $q<1$ has
been investigated. The goal of this paper is to consider the situation
when $q>1$.

As in Ref.~\cite{Ruseckas2016} we are investigating a small system
interacting with a large reservoir via short-range forces. Such description
is not directly applicable to a subsystem of a large system with long-range
interactions, where the non-extensive statistical mechanics has been
usually applied. However, the much simpler situation of short-range
forces allows us to highlight the differences from the standard statistical
mechanics and to gain a deeper insight into non-extensive statistics.

The case of $q>1$ present additional difficulties compared with the
situation when $q<1$. For example, let us consider the microcanonical
ensemble where the probability of a microstate $\mu$ is $p(\mu)=1/W$,
with $W$ being the number of microstates. If the generalized entropy
$S_{q}$ is extensive and proportional to the number of particles
$N$ in the system, the number of microstates $W$ behaves as $(1-(q-1)AN)^{-1/(q-1)}$.
Thus the number of microstates becomes infinite when the number of
particles $N$ approaches a finite maximum number $N_{\mathrm{crit}}$
and the macroscopic limit $N\rightarrow\infty$ cannot be taken. In
this situation one can try to take a different limit, $N\rightarrow N_{\mathrm{crit}}$,
instead of the limit $N\rightarrow\infty$. Additional problem is
that the $q$-exponential distributions with $q>1$ lead to divergences
in the thermodynamic limit for classical Hamiltonian systems \cite{Lutsko2011}.

The paper is organized as follows: In Section~\ref{sec:non-extensive}
we consider the canonical ensemble in the non-extensive statistical
mechanics describing a small system interacting with a large reservoir
via short-range forces. In Section~\ref{sec:divergence} we investigate
possible divergences arising in the description of the system using
canonical ensemble. We analyze properties of the generalized thermodynamical
quantities for the case of $q>1$ in Section~\ref{sec:Legendre}.
Section~\ref{sec:conclusions} summarizes our findings.

\section{Canonical ensemble in non-extensive statistical mechanics when $q>1$}

\label{sec:non-extensive}As in Ref.~\cite{Ruseckas2016} we will
consider a composite system consisting of a small system $\mathrm{S}$
interacting with a large reservoir $\mathrm{R}$. We assume that the
interaction between the system $\mathrm{S}$ and the reservoir $\mathrm{R}$
is via short-range forces, however the reservoir $\mathrm{R}$ is
not described by the Boltzmann-Gibbs statistics. We require that the
$q$-heat capacity $C_{q}^{(\mathrm{R})}$ of the reservoir, defined
by Eq.~(\ref{eq:q-heat-rez}), instead of standard heat capacity
should be large. In this article we consider only the situation when
$q>1$.

Similar investigation of the canonical ensemble in the non-extensive
statistical mechanics has been performed in Ref.~\cite{Abe2001},
however the reservoir has been considered as a heat bath. A system
weakly coupled to a finite reservoir has been considered in Ref.~\cite{Plastino1994}.
Assuming that the number of microstates of the reservoir with energy
less than $E_{\mathrm{R}}$ grows as a power-law of $E_{\mathrm{R}}$,
the $q$-exponential distribution of the energy of the system has
been obtained. In Ref.~\cite{Plastino1994} the parameter $q$ tends
to $1$ when the number of particles of the reservoir increases. Here
we do not assume any particular dependence of the parameter $q$ on
the number of particles in the reservoir.

The probability of the microstate of the system $\mathrm{S}$ can
be obtained similarly as for the case of $q<1$, considered in Ref.~\cite{Ruseckas2016}.
The total number of microstates $W_{\mathrm{tot}}(E_{\mathrm{tot}})$
of the combined system can be expressed as a sum over all available
energies of the system $\mathrm{S}$,
\begin{equation}
W_{\mathrm{tot}}(E_{\mathrm{tot}})=\sum_{E}W(E)W_{\mathrm{R}}(E_{\mathrm{tot}}-E)\,,\label{eq:num-micro-0}
\end{equation}
where $W(E)$ is the number of microstates in the system $\mathrm{S}$
having the energy $E$ and $W_{\mathrm{R}}(E_{\mathrm{R}})$ is the
number of microstates in the reservoir. Assuming that in the non-extensive
statistical mechanics the postulate of equal probabilities of microstates
in the equilibrium remains valid, the probability of the system $\mathrm{S}$
being in the microstate $\mu$ and the reservoir being in the microstate
$\mu_{\mathrm{R}}$ is equal to
\begin{equation}
p(\mu\otimes\mu_{\mathrm{R}})=\frac{1}{W(E_{\mathrm{tot}})}\,.
\end{equation}
The probability of the microstate $\mu$ of the system $\mathrm{S}$
is obtained by summing over microstates of the reservoir,
\begin{equation}
p(\mu)=\sum_{\mu_{\mathrm{R}}}p(\mu\otimes\mu_{\mathrm{R}})\,.
\end{equation}
When the energy of the microstate $\mu$ is $E_{\mu}$, the number
of possible microstates of the reservoir is $W_{\mathrm{R}}(E_{\mathrm{tot}}-E_{\mu})$
and the expression for the probability of the microstate becomes
\begin{equation}
p(\mu)=\frac{W_{\mathrm{R}}(E_{\mathrm{tot}}-E_{\mu})}{W(E_{\mathrm{tot}})}\,.\label{eq:prob-micro}
\end{equation}
As Eq.~(\ref{eq:prob-micro}) shows, from the postulate of equal
probabilities of microstates follows that the statistics of the system
$\mathrm{S}$ is determined by the reservoir. Therefore, even an ordinary
system interacting with the reservoir having large $q$-heat capacity
can be described by the $q$-entropy.

In terms of the generalized entropy of the system $S_{q}(E)=k_{\mathrm{B}}\ln_{q}W(E)$
and the generalized entropy of the reservoir $S_{q}^{(\mathrm{R})}(E_{\mathrm{R}})=k_{\mathrm{B}}\ln_{q}W_{\mathrm{R}}(E_{\mathrm{R}})$
Eq.~(\ref{eq:num-micro-0}) reads
\begin{equation}
W_{\mathrm{tot}}(E_{\mathrm{tot}})=\sum_{E}e_{q}^{\frac{1}{k_{\mathrm{B}}}S_{q}(E)}e_{q}^{\frac{1}{k_{\mathrm{B}}}S_{q}^{(\mathrm{R})}(E_{\mathrm{tot}}-E)}\,.\label{eq:num-micro-2}
\end{equation}
Differently from the case of $q<1$, this sum can be approximated
by the largest term when $q\geqslant1$. Approximation of a sum of
large $q$-exponentials is investigated in Appendix~\ref{sec:sum-exponentials}.

Since each microstate of the composite system has the same probability,
the largest term in the sum (\ref{eq:num-micro-2}) corresponds to
the most probable state of the composite system. As in Ref.~\cite{Ruseckas2016},
the condition of the maximum probability leads to the inverse temperature
\begin{equation}
\frac{1}{T}=\frac{\frac{\partial}{\partial U}S_{q}(U)}{1-\frac{q-1}{k_{\mathrm{B}}}S_{q}(U)}=\frac{\frac{\partial}{\partial E_{\mathrm{tot}}}S_{q}^{(\mathrm{R})}(E_{\mathrm{tot}}-U)}{1-\frac{q-1}{k_{\mathrm{B}}}S_{q}^{(\mathrm{R})}(E_{\mathrm{tot}}-U)}\,.\label{eq:temp-non-ext-2}
\end{equation}
where $U$ is the most-probable energy of the system. From Eq.~(\ref{eq:temp-non-ext-2})
it follows that the heat capacity of the reservoir can be expressed
as
\begin{equation}
C_{\mathrm{R}}=\frac{1}{\frac{T}{T_{q}^{(R)}}\frac{1}{C_{q}^{(\mathrm{R})}}-\frac{q-1}{k_{\mathrm{B}}}}\,,\label{eq:c-R}
\end{equation}
where
\begin{equation}
\frac{1}{T_{q}^{(\mathrm{R})}}=\frac{\partial}{\partial E_{\mathrm{R}}}S_{q}^{(\mathrm{R})}(E_{\mathrm{R}})\label{eq:q-temp-rez}
\end{equation}
is the auxiliary $q$-temperature of the reservoir and
\begin{equation}
C_{q}^{(\mathrm{R})}=-\frac{1}{(T_{q}^{(\mathrm{R})})^{2}\frac{\partial^{2}}{\partial E_{\mathrm{R}}^{2}}S_{q}^{(\mathrm{R})}(E_{\mathrm{R}})}\label{eq:q-heat-rez}
\end{equation}
is the $q$-heat capacity of the reservoir, defined similarly to the
heat capacity in standard statistical mechanics. Equation similar
to Eq.~(\ref{eq:c-R}) has been obtained in Ref.~\cite{Wada2002}.
In the formulation of the non-extensive statistical mechanics based
on maximization of entropy, the auxiliary temperature $T_{q}$ appears
as the inverse of the Lagrange multiplier associated with the energy
constraint. This temperature can have a physical meaning in systems
with long-range interactions. For example, temperature $T_{q}$ is
related to the density of vortices in type II superconductors \cite{Nobre2015}. 

If we introduce the entropy of the combined system as $S_{q}^{(\mathrm{tot})}(E_{\mathrm{tot}})=k_{\mathrm{B}}\ln_{q}W(E_{\mathrm{tot}})$
then approximating the sum (\ref{eq:num-micro-2}) by the largest
term we get that the entropy of the combined system is a pseudo-additive
combination of the entropies of the system $\mathrm{S}$ and the reservoir
$\mathrm{R}$: 
\begin{equation}
S_{q}^{(\mathrm{tot})}(E_{\mathrm{tot}})\approx S_{q}(U)+S_{q}^{(\mathrm{R})}(E_{\mathrm{tot}}-U)-\frac{q-1}{k_{\mathrm{B}}}S_{q}(U)S_{q}^{(\mathrm{R})}(E_{\mathrm{tot}}-U)\,.
\end{equation}
According to Eq.~(\ref{eq:temp-non-ext-2}), in the ensemble considered
in this section the physical temperatures of the system and the reservoir
are equal, whereas the corresponding $q$-temperatures are not. If
one requires equality of $q$-temperature, the additivity of energies
does not apply \cite{Ou2006}. However, when the interactions between
the system $\mathrm{S}$ and $\mathrm{R}$ are long range and, consequently,
the energy is not additive, then the pseudo-additivity of entropies
together with equality of $q$-temperatures can be valid \cite{Scarfone2010}.

Similarly as in Ref.~\cite{Ruseckas2016} for the $q<1$ case, we
assume that the second derivative of $q$-entropy of the reservoir
is very small, $\frac{\partial^{2}}{\partial E_{\mathrm{tot}}^{2}}S_{q}^{(\mathrm{R})}(E_{\mathrm{tot}})\approx0$,
and, consequently, the $q$-heat capacity of the reservoir, defined
by Eq.~(\ref{eq:q-heat-rez}), is very large. Taking the limit $C_{q}^{(\mathrm{R})}\rightarrow\infty$
in Eq.~(\ref{eq:c-R}) we obtain the heat capacity of the reservoir
$C_{\mathrm{R}}=-\frac{k_{\mathrm{B}}}{q-1}$. The heat capacity is
\emph{negative} when $q>1$. Increase of the energy of the reservoir
with very large $q$-heat capacity by $\Delta E$ leads to the new
temperature of the reservoir 
\begin{equation}
T^{\prime}=T-\frac{q-1}{k_{\mathrm{B}}}\Delta E\,.\label{eq:t-shift}
\end{equation}
The temperature of the reservoir \emph{decreases} by increasing the
energy. 

Possibility of negative heat capacity in the case of $q>1$ has been
implied in Ref.~\cite{Wada2002}. Such a system is thermodynamically
unstable. However, we want to point out that heat capacity is not
always positive in microcanonical ensemble theory \cite{Thirring1983,Posch1990,Posch1993,Latora1998,Antoni1998}.
Specifically, this is the case for systems interacting through long-range
forces \cite{Dauxois2002}, where nonequivalence of the microcanonical
and canonical ensembles \cite{Barre2001,Mukamel2005} and negative
microcanonical specific heat \cite{Lynden-Bell1968,Thirring1970}
has been demonstrated. Specific heat can take a negative value in
thermodynamics of a self-gravitating system \cite{Saslaw1985}, leading
to the so-called gravithermal instability. Negative heat capacities
have been predicted for melting atomic clusters \cite{Bixon1989,Labastie1990}
and fragmenting nuclei \cite{Gross1990}. Experimentally negative
heat capacity has been observed in excited nuclear systems \cite{DAgostino2000}
and in sodium clusters \cite{Schmidt2001}. In the context of non-extensive
statistical mechanics, negative specific heat has been obtained for
$q$-ideal gas \cite{Abe1999a,Abe2000} and for two-level systems
\cite{DiSisto1999}.

Assuming very small second derivative of $q$-entropy of the reservoir,
the number of microstates of the reservoir can be approximated as
\begin{equation}
W_{\mathrm{R}}(E_{\mathrm{tot}}-E)=e_{q}^{\frac{1}{k_{\mathrm{B}}}S_{q}^{(\mathrm{R})}(E_{\mathrm{tot}}-E)}\approx e_{q}^{\frac{1}{k_{\mathrm{B}}}S_{q}^{(\mathrm{R})}(E_{\mathrm{tot}}-U)-\frac{1}{k_{\mathrm{B}}}(E-U)\frac{\partial}{\partial E_{\mathrm{tot}}}S_{q}^{(\mathrm{R})}(E_{\mathrm{tot}}-U)}\,.\label{eq:q-expansion}
\end{equation}
Using Eqs.~(\ref{eq:prob-micro}) and (\ref{eq:q-expansion}) we
obtain that the probability of the microstate of the system $\mathrm{S}$
is proportional to the factor
\begin{equation}
\tilde{P}(E)=\exp_{q}\left(-\frac{1}{k_{\mathrm{B}}T(U)}(E-U)\right)\,,\label{eq:factor-non-ext-0}
\end{equation}
where the temperature $T(U)$ is given by Eq.~(\ref{eq:temp-non-ext-2}).
In contrast to the thermostat with the very large heat capacity, the
temperature $T(U)$ depends not only on the reservoir but also on
the properties of the system. Therefore, it is convenient to introduce
the temperature of the isolated reservoir
\begin{equation}
\frac{1}{T(0)}=\frac{\frac{\partial}{\partial E_{\mathrm{tot}}}S_{q}^{(\mathrm{R})}(E_{\mathrm{tot}})}{1-\frac{q-1}{k_{\mathrm{B}}}S_{q}^{(\mathrm{R})}(E_{\mathrm{tot}})}\,.\label{eq:temp-zero}
\end{equation}
Using Eqs.~(\ref{eq:temp-non-ext-2}), (\ref{eq:temp-zero}) together
with the assumption $\frac{\partial^{2}}{\partial E_{\mathrm{R}}^{2}}S_{q}^{(\mathrm{R})}(E_{\mathrm{R}})\approx0$
we get
\begin{equation}
T(U)\approx T(0)+\frac{q-1}{k_{\mathrm{B}}}U\,.\label{eq:tu-t0}
\end{equation}
This equation shows that the interaction with the system \emph{raises}
the temperature of the reservoir. However, due to the large $q$-heat
capacity the $q$-temperature of the reservoir, defined by Eq.~(\ref{eq:q-temp-rez}),
remains constant. Inserting Eq.~(\ref{eq:tu-t0}) into Eq.~(\ref{eq:factor-non-ext-0})
we get that the probability of the microstate of the system $\mathrm{S}$
is proportional to the factor
\begin{equation}
P(E)=\exp_{q}\left(-\frac{1}{k_{\mathrm{B}}T(0)}E\right)\,.\label{eq:factor-non-ext}
\end{equation}
An expression similar to Eq.~(\ref{eq:factor-non-ext}) has been
obtained in Ref.~\cite{Abe2001}. Using the factor (\ref{eq:factor-non-ext})
we can write the normalized probability of the microstate as
\begin{equation}
p(\mu)=\frac{1}{Z_{q}}e_{q}^{-\frac{1}{k_{\mathrm{B}}T(0)}E_{\mu}}\,,\label{eq:q-prob-normalized}
\end{equation}
where
\begin{equation}
Z_{q}=\sum_{\mu}e_{q}^{-\frac{1}{k_{\mathrm{B}}T(0)}E_{\mu}}\label{eq:zq-1}
\end{equation}
is the generalized partition function.

On the first sight the factor (\ref{eq:factor-non-ext}) is not invariant
to the change of zero of energies. However, as in Ref.~\cite{Ruseckas2016},
we can argue that the shift of the energy zero of the system by $\Delta E$
is equivalent to the decrease of the energy of the reservoir by $\Delta E$
leading to the increase of the temperature. From the requirement that
the probability of the microstate should remain the same follows that
the new factor should be proportional to the old, 
\begin{equation}
P^{\prime}(E)=\exp_{q}\left(-\frac{1}{k_{\mathrm{B}}T^{\prime}(0)}E\right)\sim P(E+\Delta E)=\exp_{q}\left(-\frac{1}{k_{\mathrm{B}}T(0)}(E+\Delta E)\right)\,.
\end{equation}
Consequently, the new temperature of the reservoir should be equal
to
\begin{equation}
T^{\prime}(0)=T(0)+\frac{q-1}{k_{\mathrm{B}}}\Delta E\,.
\end{equation}
This equation is consistent with Eq.~(\ref{eq:t-shift}).

\section{Divergences in canonical ensemble approach}

\label{sec:divergence}In the canonical ensemble approach the description
of the reservoir is simplified to just one number, the temperature.
The validity of such a simplification depends the system interacting
with the reservoir. Namely, it is assumed that the system should be
much smaller that the reservoir; the precise requirement depends on
statistics. Let us consider the standard, Boltzmann-Gibbs statistical
mechanics at first. In the derivation of the Boltzmann factor an assumption
is made that the number $W(E)$ of microstates of the system having
energy $E_{\mu}=E$ should not grow fast with increasing energy and
the distribution of the energy $p(E)$ should be normalizable, 
\begin{equation}
\int W(E)\exp\left(-\frac{1}{k_{\mathrm{B}}T}E\right)dE<\infty\,.
\end{equation}
For a hypothetical system where the number of microstates $W(E)$
grows with increasing energy as fast as $E^{-1}e^{\frac{1}{k_{\mathrm{B}}T}E}$
or faster, this assumption is not satisfied and the reservoir cannot
be considered as a thermostat. Such a system is not smaller than the
reservoir. In order to get normalizable probabilities in this situation
one should consider the reservoir as a finite system having finite
energy. Thus the canonical ensemble leading to the exponential Boltzmann
factor is not applicable when the number of microstates grow exponentially.

Now let us examine the situation described in the previous Section,
when the $q$-heat capacity of the reservoir is large when $q>1$.
Similarly as in the Boltzmann-Gibbs statistical mechanics the description
using canonical ensemble can be applied only when the system is small
and the number $W(E)$ of microstates having energy $E_{\mu}=E$ grows
with increasing energy slow enough. Using the factor (\ref{eq:factor-non-ext})
we get that probability is normalizable when $W(E)$ grows with increasing
energy slower than $E^{\frac{1}{q-1}-1}$. That is, at large energies
$W(E)$ should grow slower than $\exp_{q_{\mathrm{lim}}}(aE)$ with
\begin{equation}
q_{\mathrm{lim}}=2-\frac{1}{2-q}\,.\label{eq:q-large}
\end{equation}
When $q=1$ we get $q_{\mathrm{lim}}=1$, which coincides with the
limit on the growth in the Boltzmann-Gibbs statistical mechanics.
According to Eq.~(\ref{eq:q-large}), $q_{\mathrm{lim}}<1$ when
$1<q<2$. If the number of microstates of the system $W(E)$ grows
with increasing energy faster than this limit then to get finite probabilities
the reservoir should be described as a finite system having finite
energy and the generalized canonical ensemble is not applicable. When
$q>1$, this situation can occur for conventional physical systems,
e.g. for classical Hamiltonian systems in the thermodynamic limit
\cite{Lutsko2011}. This problem has been first noticed by Abe \cite{Abe1999a}
by trying to describe ideal gas where the effects of the interaction
are replaced by the introduction of $q\neq1$.

Similar limitation occurs also in the case of $q<1$. Since the distribution
of energies when $q<1$ has a cut off $E_{\mathrm{max}}$, this allows
for the number of microstates of the system to grow with increasing
energy even faster than in the case of Boltzmann-Gibbs canonical ensemble.
However, if the number of microstates is singular when $E$ approaches
$E_{\mathrm{max}}$, the probability can become unnormalizable. To
get finite probabilities the number of microstates $W(E)$ when energy
approaches $E_{\mathrm{max}}$ should grow slower than $\exp_{q_{\mathrm{lim}}}(aE)$,
where $a=1/[(q_{\mathrm{lim}}-1)E_{\mathrm{max}}]$. Here the value
of $q_{\mathrm{lim}}$ is given by the same equation (\ref{eq:q-large}).
Thus we can conclude that for all possible values of $q$ the number
of microstates $W(E)$ should grow with increasing energy slower that
the $q$-exponential with the limiting value of $q$ (\ref{eq:q-large}).

The simplest way to take into account the finiteness of the reservoir
is to introduce a cut-off energy $E_{\mathrm{max}}$ into the probability
of the microstate:
\begin{equation}
p^{\prime}(\mu)=\frac{1}{Z_{q}^{\prime}}e_{q}^{-\frac{1}{k_{\mathrm{B}}T(0)}E_{\mu}}\Theta(E_{\mathrm{max}}-E_{\mu})\,.
\end{equation}
Here $\Theta$ is the Heaviside step function. The cut-off energy
$E_{\mathrm{max}}$ has the meaning of the finite energy of the isolated
reservoir. Similar possibility has been suggested in Ref.~\cite{Lutsko2011}.
The cut-off using the step function is only the simplest possibility,
the specific form of the cut-off depends on the details of the reservoir.
Such a description is outside of the formalism of canonical ensemble
where reservoir is characterized only by temperature.

In Ref.~\cite{Plastino2013} it was suggested to remove the divergences
occurring in the case of $q>1$ by calculating the $q$-partition
function $Z_{q}$ as a $q$-Laplace transform of the energy density.
However, this proposal is problematic, as it is pointed out in Ref.~\cite{Lutsko2014},
because the introduction of the $q$-Laplace transform only removes
divergences in the averages of the functions of energy.

Note, that we obtained the non-applicability of the canonical ensemble
in the non-extensive statistical mechanics for the systems where the
growth of the number of microstates with the energy is faster than
$q$-exponential with $q=q_{\mathrm{lim}}$ using the assumption of
short-range interactions between the system and and the reservoir.
In the case of long-range interactions this result is not necessarily
valid. When interactions are long-range, the systems can be non-ergodic
and not all available microstates can be reached. In this situation
the description using microcanonical ensemble should be modified,
for example, assigning equal probabilities only to reachable microstates.
The effective number of reachable microstates can grow slower than
in the ergodic case and the probability proportional to the $q$-expoential
with $q>1$ can be applicable.

\section{Generalized thermodynamical quantities}

\label{sec:Legendre}As for the case of $q<1$, considered in Ref.~\cite{Ruseckas2016},
there are several different possibilities to generalize the free energy.
All equations of Ref.~\cite{Ruseckas2016} where no approximations
have been made remain valid also for $q>1$. In this Section we highlight
only the differences between $q>1$ and $q<1$ cases.

Let us consider the generalized partition function $Z_{q}$, given
by Eq.~(\ref{eq:zq-1}). The distribution of the energy of the system
$E$ is equal to the probability $p(\mu)$ multiplied by the number
$W(E)=e_{q}^{\frac{1}{k_{\mathrm{B}}}S_{q}(E)}$ of microstates having
energy $E_{\mu}=E$. Thus the generalized partition function $Z_{q}$
can be written as a sum over energies
\begin{equation}
Z_{q}=\sum_{E}e_{q}^{\frac{1}{k_{\mathrm{B}}}S_{q}(E)}e_{q}^{-\frac{1}{k_{\mathrm{B}}T(0)}E}=\sum_{E}e_{q}^{\frac{1}{k_{\mathrm{B}}}\frac{T(E)}{T(0)}S_{q}(E)-\frac{1}{k_{\mathrm{B}}T(0)}E}\,,\label{eq:zq-e}
\end{equation}
where
\begin{equation}
T(E)=T(0)+\frac{q-1}{k_{\mathrm{B}}}E\,.
\end{equation}
From the properties of $q$-logarithm (\ref{eq:q-log1}) with $q>1$
follows that the entropy $S_{q}(E)$ is smaller than the $q$-dependent
maximum value,
\begin{equation}
S_{q}(E)<\frac{k_{\mathrm{B}}}{q-1}\,.
\end{equation}
When $q>1$ and the entropy $S_{q}(U)$ corresponding to the most-probable
energy of the system $U$ is close to the limiting value $k_{\mathrm{B}}/(q-1)$,
the sum in Eq.~(\ref{eq:zq-e}) can be approximated by the largest
term. The approximation of the sum of large $q$-exponentials with
$q>1$ is investigated in Appendix~\ref{sec:sum-exponentials}. Thus
the $q$-logarithm of the sum in Eq.~(\ref{eq:zq-e}) can be approximated
as
\begin{equation}
\ln_{q}Z_{q}\approx\frac{1}{k_{\mathrm{B}}}S_{q}(U)-\frac{1-\frac{q-1}{k_{\mathrm{B}}}S_{q}(U)}{k_{\mathrm{B}}T(0)}U\,.\label{eq:lnqzq-appr}
\end{equation}
The unnormalized $q$-average energy of the system
\begin{equation}
\bar{U}_{q}=\sum_{\mu}E_{\mu}p(\mu)^{q}
\end{equation}
can be calculated using the equation \cite{Ruseckas2016}
\begin{equation}
\bar{U}_{q}=k_{\mathrm{B}}T(0)^{2}\frac{\partial}{\partial T(0)}\ln_{q}Z_{q}\,.\label{eq:u1}
\end{equation}
From the approximation (\ref{eq:lnqzq-appr}) we get
\begin{equation}
\bar{U}_{q}\approx\left(1-\frac{q-1}{k_{\mathrm{B}}}S_{q}(U)\right)U\,.\label{eq:uq-appr}
\end{equation}
As have been shown in Ref.~\cite{Ruseckas2016}, the entropy
\begin{equation}
\bar{S}_{q}=k_{\mathrm{B}}\frac{1-\sum_{\mu}p(\mu)^{q}}{q-1}\,.\label{eq:q-entr-2}
\end{equation}
can be obtained using the equation
\begin{equation}
\bar{F}_{q}=\bar{U}_{q}-T(0)\bar{S}_{q}\,,\label{eq:fq1-2}
\end{equation}
where 
\begin{equation}
\bar{F}_{q}=-k_{\mathrm{B}}T(0)\ln_{q}Z_{q}\,.\label{eq:fq1}
\end{equation}
is the generalized free energy corresponding to the temperature $T(0)$.
Using the approximation (\ref{eq:lnqzq-appr}) and Eqs.~(\ref{eq:uq-appr}),
(\ref{eq:fq1-2}) we obtain 
\[
\bar{S}_{q}\approx S_{q}(U)\,.
\]
Thus, similarly as in Boltzmann-Gibbs statistics and differently than
in the case with $q<1$, the average entropy $\bar{S}_{q}$ for $q>1$
is approximately equal to the maximal entropy $S_{q}(U)$. The approximation
$\bar{S}_{q}\approx S_{q}(U)$ is consistent with Eq.~(\ref{eq:q-entr-2}).
Indeed, we have
\begin{equation}
\bar{S}_{q}=k_{\mathrm{B}}\frac{1-\sum_{\mu}p(\mu)^{q}}{q-1}=k_{\mathrm{B}}\frac{1-\sum_{E}\left(1-\frac{q-1}{k_{\mathrm{B}}}S_{q}(E)\right)p(E)^{q}}{q-1}\,.
\end{equation}
When the maximum of the entropy $S_{q}(U)$ is close to the limiting
value, approximating the sum by the largest term corresponding to
$E=U$ we get $\bar{S}_{q}\approx S_{q}(U)$.

The normalized $q$-average of the energy
\begin{equation}
U_{q}=\frac{\sum_{\mu}E_{\mu}p(\mu)^{q}}{\sum_{\mu}p(\mu)^{q}}
\end{equation}
is related to the unnormalized $q$-average as \cite{Ruseckas2016}
\begin{equation}
U_{q}=\frac{\bar{U}_{q}}{1-\frac{q-1}{k_{\mathrm{B}}}\bar{S}_{q}}\,.\label{eq:uq-norm-unnorm}
\end{equation}
Using the approximation (\ref{eq:uq-appr}) we obtain
\begin{equation}
U_{q}\approx U\,.
\end{equation}
Differentiating the expression for the average energy of the system
\begin{equation}
\bar{U}=\sum_{\mu}E_{\mu}p(\mu)\label{eq:u-aver}
\end{equation}
with respect to the temperature $T(0)$ and using Eqs.~(\ref{eq:q-prob-normalized}),
(\ref{eq:u1}), and (\ref{eq:uq-norm-unnorm}) we can express the
difference between the average energy and normalized $q$-average
energy as
\begin{equation}
\bar{U}-U_{q}=(q-1)T(U_{q})\frac{\partial}{\partial T(0)}\bar{U}\,.\label{eq:u-avg-diff}
\end{equation}
As this equation shows, when $q\neq1$ the difference between different
averages is proportional to the physical temperature $T(U_{q})$.
When $q>1$ and the maximum of the entropy $S_{q}(U)$ is close to
the limiting value, the probability of the energy $E=U$ is much larger
than the probabilities of other energy values. In this case $\bar{U}\approx U_{q}$
and from Eq.~(\ref{eq:u-avg-diff}) follows that $\partial\bar{U}/\partial T(0)\approx0$.

The auxiliary $q$-temperature $T_{q}$ of the system $\mathrm{S}$,
defined as 
\begin{equation}
\frac{1}{T_{q}}=\frac{\partial\bar{S}_{q}}{\partial U_{q}}\label{eq:tq-tilde}
\end{equation}
is related via the equation
\begin{equation}
T(U_{q})=T_{q}\left(1-\frac{q-1}{k_{\mathrm{B}}}\bar{S}_{q}\right)\label{eq:tq-tilde-t}
\end{equation}
to the temperature $T(U_{q})=T(0)+\frac{q-1}{k_{\mathrm{B}}}U_{q}$
of the reservoir corresponding to the energy of the system equal to
$U_{q}$ \cite{Ruseckas2016}. Since $\bar{S}_{q}>0$, the $q$-temperature
is always larger than the physical temperature $T(U_{q})$. In contrast,
the $q$-temperature is smaller than the physical temperature when
$q<1$.

The physical heat capacity $C$, obtained as the derivative of $U_{q}$
with respect to the physical temperature $T(U_{q})$, 
\begin{equation}
C=\frac{\partial U_{q}}{\partial T(U_{q})}
\end{equation}
is related to the $q$-heat capacity of the system
\begin{equation}
C_{q}=\frac{\partial U_{q}}{\partial T_{q}}=T_{q}\frac{\partial\bar{S}_{q}}{\partial T_{q}}
\end{equation}
via the equation \cite{Ruseckas2016} 
\begin{equation}
C=\frac{1}{\frac{T(U_{q})}{T_{q}}\frac{1}{C_{q}}-\frac{q-1}{k_{\mathrm{B}}}}\,.\label{eq:c-cq-tilde}
\end{equation}
Similar equation has been obtained in~Ref.~\cite{Wada2002}. Since
$T_{q}>T(U_{q})$ when $q>1$, from Eq.~(\ref{eq:c-cq-tilde}) follows
that the physical heat capacity $C$ is always larger than the $q$-heat
capacity $C_{q}$. In contrast, for $q<1$ the physical heat capacity
$C$ is always smaller than the $q$-heat capacity $C_{q}$.

\section{Conclusions}

\label{sec:conclusions}In summary, we have considered a small system
interacting via short-range forces with a large reservoir that has
large $q$-heat capacity with $q>1$. Such a system can be described
by the non-extensive statistical mechanics, with the probability of
the microstate of the system given by the $q$-exponential (\ref{eq:factor-non-ext})
instead of the usual Boltzmann factor. The reservoir can be described
using the generalized entropy and exhibit large $q$-heat capacity
only when long-range interactions and long-range correlations are
present. Since we assumed short-range interactions of the system under
consideration with the reservoir, the approach presented in this paper
is not applicable to a subsystem of such a reservoir.

The assumption of large $q$-heat capacity leads to a negative physical
heat capacity, thus the description using canonical ensemble with
$q>1$ is applicable only when the system is interacting with negative
heat capacity reservoir. Although negative heat capacity means thermodynamical
instability, systems with long-range interactions can exhibit negative
microcanonical specific heat \cite{Dauxois2002}. Due to finite heat
capacity of the reservoir the physical temperature in the equilibrium
$T$ depends both on the properties of the reservoir and the properties
of the system. On the other hand, the auxiliary $q$-temperature $T_{q}^{(\mathrm{R})}$
(\ref{eq:q-temp-rez}) remains constant due to large $q$-heat capacity
of the reservoir.

The requirement that the system interacting with the reservoir should
be small limits the growth of the number of microstates of the system
$W(E)$ with increasing energy. We obtained that the description using
the canonical ensemble is applicable only when $W(E)$ grows slower
than $q$-exponential with the value of $q$ given by Eq.~(\ref{eq:q-large}).
This limit is valid for all values of $q$, for $q>1$ as well as
$q=1$ and $q<1$.

\appendix

\section{Sum of large $q$-exponentials}

\label{sec:sum-exponentials}

Let us consider the sum of large $q$-exponentials
\begin{equation}
Z_{q}=\sum_{i=1}^{W}e_{q}^{N\phi(i)}\label{eq:app-sum}
\end{equation}
with $q>1$. There is a maximum $N=N_{\mathrm{crit}}$ when one of
the terms becomes infinite. The limiting value $N_{\mathrm{crit}}$
is determined from the condition $(q-1)N_{\mathrm{crit}}\phi_{\mathrm{max}}=1$,
where $\phi_{\mathrm{max}}$ is the maximum of $\phi(i)$. When $N$
is close to $N_{\mathrm{crit}}$ then the sum of large $q$-exponentials
with $q>1$ can be approximated by the largest term. Since $\phi_{\mathrm{max}}$
is the maximum of $\phi(i)$, the sum $Z_{q}$ satisfies the following
inequality:
\begin{equation}
e_{q}^{N\phi_{\mathrm{max}}}\leqslant Z_{q}\leqslant We_{q}^{N\phi_{\mathrm{max}}}\,.
\end{equation}
From this inequality follows that
\begin{equation}
0\leqslant\frac{\ln_{q}Z_{q}}{N}-\phi_{\mathrm{max}}\leqslant\frac{\ln_{q}W}{N}-(q-1)\phi_{\mathrm{max}}\ln_{q}W\,.
\end{equation}
Inserting
\begin{equation}
\phi_{\mathrm{max}}=\frac{1}{(q-1)N_{\mathrm{crit}}}
\end{equation}
we obtain
\begin{equation}
0\leqslant\frac{\ln_{q}Z_{q}}{N}-\phi_{\mathrm{max}}\leqslant\frac{\ln_{q}W}{N}\left(1-\frac{N}{N_{\mathrm{crit}}}\right)\,.
\end{equation}
In the limit $N\rightarrow N_{\mathrm{crit}}$ the multiplier $1-N/N_{\mathrm{crit}}$
vanishes, therefore 
\begin{equation}
\lim_{N\rightarrow N_{\mathrm{crit}}}\frac{\ln_{q}Z_{q}}{N}=\phi_{\mathrm{max}}\,.
\end{equation}
This limit shows that the sum of $q$-exponentials (\ref{eq:app-sum})
with $q>1$ can be approximated by the largest term when $N$ is close
to $N_{\mathrm{crit}}$.


\begin{thebibliography}{47}%
\makeatletter
\providecommand \@ifxundefined [1]{%
 \@ifx{#1\undefined}
}%
\providecommand \@ifnum [1]{%
 \ifnum #1\expandafter \@firstoftwo
 \else \expandafter \@secondoftwo
 \fi
}%
\providecommand \@ifx [1]{%
 \ifx #1\expandafter \@firstoftwo
 \else \expandafter \@secondoftwo
 \fi
}%
\providecommand \natexlab [1]{#1}%
\providecommand \enquote  [1]{``#1''}%
\providecommand \bibnamefont  [1]{#1}%
\providecommand \bibfnamefont [1]{#1}%
\providecommand \citenamefont [1]{#1}%
\providecommand \href@noop [0]{\@secondoftwo}%
\providecommand \href [0]{\begingroup \@sanitize@url \@href}%
\providecommand \@href[1]{\@@startlink{#1}\@@href}%
\providecommand \@@href[1]{\endgroup#1\@@endlink}%
\providecommand \@sanitize@url [0]{\catcode `\\12\catcode `\$12\catcode
  `\&12\catcode `\#12\catcode `\^12\catcode `\_12\catcode `\%12\relax}%
\providecommand \@@startlink[1]{}%
\providecommand \@@endlink[0]{}%
\providecommand \url  [0]{\begingroup\@sanitize@url \@url }%
\providecommand \@url [1]{\endgroup\@href {#1}{\urlprefix }}%
\providecommand \urlprefix  [0]{URL }%
\providecommand \Eprint [0]{\href }%
\providecommand \doibase [0]{http://dx.doi.org/}%
\providecommand \selectlanguage [0]{\@gobble}%
\providecommand \bibinfo  [0]{\@secondoftwo}%
\providecommand \bibfield  [0]{\@secondoftwo}%
\providecommand \translation [1]{[#1]}%
\providecommand \BibitemOpen [0]{}%
\providecommand \bibitemStop [0]{}%
\providecommand \bibitemNoStop [0]{.\EOS\space}%
\providecommand \EOS [0]{\spacefactor3000\relax}%
\providecommand \BibitemShut  [1]{\csname bibitem#1\endcsname}%
\let\auto@bib@innerbib\@empty
\bibitem [{\citenamefont {Tsallis}(2009{\natexlab{a}})}]{Tsallis2009-1}%
  \BibitemOpen
  \bibfield  {author} {\bibinfo {author} {\bibfnamefont {C.}~\bibnamefont
  {Tsallis}},\ }\href@noop {} {\emph {\bibinfo {title} {Introduction to
  Nonextensive Statistical Mechanics---Approaching a Complex World}}}\
  (\bibinfo  {publisher} {Springer},\ \bibinfo {address} {New York},\ \bibinfo
  {year} {2009})\BibitemShut {NoStop}%
\bibitem [{\citenamefont {Liu}\ and\ \citenamefont {Goree}(2008)}]{Liu2008}%
  \BibitemOpen
  \bibfield  {author} {\bibinfo {author} {\bibfnamefont {B.}~\bibnamefont
  {Liu}}\ and\ \bibinfo {author} {\bibfnamefont {J.}~\bibnamefont {Goree}},\
  }\href@noop {} {\bibfield  {journal} {\bibinfo  {journal} {Phys. Rev. Lett.}\
  }\textbf {\bibinfo {volume} {100}},\ \bibinfo {pages} {055003} (\bibinfo
  {year} {2008})}\BibitemShut {NoStop}%
\bibitem [{\citenamefont {DeVoe}(2009)}]{DeVoe2009}%
  \BibitemOpen
  \bibfield  {author} {\bibinfo {author} {\bibfnamefont {R.~G.}\ \bibnamefont
  {DeVoe}},\ }\href@noop {} {\bibfield  {journal} {\bibinfo  {journal} {Phys.
  Rev. Lett.}\ }\textbf {\bibinfo {volume} {102}},\ \bibinfo {pages} {063001}
  (\bibinfo {year} {2009})}\BibitemShut {NoStop}%
\bibitem [{\citenamefont {Pickup}\ \emph {et~al.}(2009)\citenamefont {Pickup},
  \citenamefont {Cywinski}, \citenamefont {Pappas}, \citenamefont {Farago},\
  and\ \citenamefont {Fouquet}}]{Pickup2009}%
  \BibitemOpen
  \bibfield  {author} {\bibinfo {author} {\bibfnamefont {R.~M.}\ \bibnamefont
  {Pickup}}, \bibinfo {author} {\bibfnamefont {R.}~\bibnamefont {Cywinski}},
  \bibinfo {author} {\bibfnamefont {C.}~\bibnamefont {Pappas}}, \bibinfo
  {author} {\bibfnamefont {B.}~\bibnamefont {Farago}}, \ and\ \bibinfo {author}
  {\bibfnamefont {P.}~\bibnamefont {Fouquet}},\ }\href@noop {} {\bibfield
  {journal} {\bibinfo  {journal} {Phys. Rev. Lett.}\ }\textbf {\bibinfo
  {volume} {102}},\ \bibinfo {pages} {097202} (\bibinfo {year}
  {2009})}\BibitemShut {NoStop}%
\bibitem [{\citenamefont {Huang}\ \emph {et~al.}(2010)\citenamefont {Huang},
  \citenamefont {Su}, \citenamefont {{El Kaabouchi}}, \citenamefont {Wang},\
  and\ \citenamefont {Chen}}]{Huang2010}%
  \BibitemOpen
  \bibfield  {author} {\bibinfo {author} {\bibfnamefont {Z.}~\bibnamefont
  {Huang}}, \bibinfo {author} {\bibfnamefont {G.}~\bibnamefont {Su}}, \bibinfo
  {author} {\bibfnamefont {A.}~\bibnamefont {{El Kaabouchi}}}, \bibinfo
  {author} {\bibfnamefont {Q.~A.}\ \bibnamefont {Wang}}, \ and\ \bibinfo
  {author} {\bibfnamefont {J.}~\bibnamefont {Chen}},\ }\href@noop {} {\bibfield
   {journal} {\bibinfo  {journal} {J. Stat. Mech.}\ }\textbf {\bibinfo {volume}
  {2010}},\ \bibinfo {pages} {L05001} (\bibinfo {year} {2010})}\BibitemShut
  {NoStop}%
\bibitem [{\citenamefont {Prehl}\ \emph {et~al.}(2012)\citenamefont {Prehl},
  \citenamefont {Essex},\ and\ \citenamefont {Hoffman}}]{Prehl2012}%
  \BibitemOpen
  \bibfield  {author} {\bibinfo {author} {\bibfnamefont {J.}~\bibnamefont
  {Prehl}}, \bibinfo {author} {\bibfnamefont {C.}~\bibnamefont {Essex}}, \ and\
  \bibinfo {author} {\bibfnamefont {K.~H.}\ \bibnamefont {Hoffman}},\
  }\href@noop {} {\bibfield  {journal} {\bibinfo  {journal} {Entropy}\ }\textbf
  {\bibinfo {volume} {14}},\ \bibinfo {pages} {701} (\bibinfo {year}
  {2012})}\BibitemShut {NoStop}%
\bibitem [{\citenamefont {Adare}\ \emph {et~al.}(2011)\citenamefont {Adare}
  \emph {et~al.}}]{Adare2011}%
  \BibitemOpen
  \bibfield  {author} {\bibinfo {author} {\bibfnamefont {A.}~\bibnamefont
  {Adare}} \emph {et~al.},\ }\href@noop {} {\bibfield  {journal} {\bibinfo
  {journal} {Phys. Rev. D}\ }\textbf {\bibinfo {volume} {83}},\ \bibinfo
  {pages} {052004} (\bibinfo {year} {2011})}\BibitemShut {NoStop}%
\bibitem [{\citenamefont {Budini}(2012)}]{Budini2012}%
  \BibitemOpen
  \bibfield  {author} {\bibinfo {author} {\bibfnamefont {A.~A.}\ \bibnamefont
  {Budini}},\ }\href@noop {} {\bibfield  {journal} {\bibinfo  {journal} {Phys.
  Rev. E}\ }\textbf {\bibinfo {volume} {86}},\ \bibinfo {pages} {011109}
  (\bibinfo {year} {2012})}\BibitemShut {NoStop}%
\bibitem [{\citenamefont {Du}(2012)}]{Du2012}%
  \BibitemOpen
  \bibfield  {author} {\bibinfo {author} {\bibfnamefont {J.-L.}\ \bibnamefont
  {Du}},\ }\href@noop {} {\bibfield  {journal} {\bibinfo  {journal} {J. Stat.
  Mech.}\ }\textbf {\bibinfo {volume} {2012}},\ \bibinfo {pages} {P02006}
  (\bibinfo {year} {2012})}\BibitemShut {NoStop}%
\bibitem [{\citenamefont {Lutz}\ and\ \citenamefont
  {Renzoni}(2013)}]{Lutz2013}%
  \BibitemOpen
  \bibfield  {author} {\bibinfo {author} {\bibfnamefont {E.}~\bibnamefont
  {Lutz}}\ and\ \bibinfo {author} {\bibfnamefont {F.}~\bibnamefont {Renzoni}},\
  }\href@noop {} {\bibfield  {journal} {\bibinfo  {journal} {Nature Physics}\
  }\textbf {\bibinfo {volume} {9}},\ \bibinfo {pages} {615} (\bibinfo {year}
  {2013})}\BibitemShut {NoStop}%
\bibitem [{\citenamefont {Beck}\ and\ \citenamefont {Miah}(2013)}]{Beck2013}%
  \BibitemOpen
  \bibfield  {author} {\bibinfo {author} {\bibfnamefont {C.}~\bibnamefont
  {Beck}}\ and\ \bibinfo {author} {\bibfnamefont {S.}~\bibnamefont {Miah}},\
  }\href@noop {} {\bibfield  {journal} {\bibinfo  {journal} {Phys. Rev. E}\
  }\textbf {\bibinfo {volume} {87}},\ \bibinfo {pages} {031002} (\bibinfo
  {year} {2013})}\BibitemShut {NoStop}%
\bibitem [{\citenamefont {Gell-Mann}\ and\ \citenamefont
  {Tsallis}(2004)}]{Gell-Mann2004}%
  \BibitemOpen
  \bibfield  {author} {\bibinfo {author} {\bibfnamefont {C.~M.}\ \bibnamefont
  {Gell-Mann}}\ and\ \bibinfo {author} {\bibfnamefont {C.}~\bibnamefont
  {Tsallis}},\ }\href@noop {} {\emph {\bibinfo {title} {Nonextensive
  Entropy---Interdisciplinary Applications}}}\ (\bibinfo  {publisher} {Oxford
  University Press},\ \bibinfo {address} {New York},\ \bibinfo {year}
  {2004})\BibitemShut {NoStop}%
\bibitem [{\citenamefont {Abe}(2006)}]{Abe2006}%
  \BibitemOpen
  \bibfield  {author} {\bibinfo {author} {\bibfnamefont {S.}~\bibnamefont
  {Abe}},\ }\href@noop {} {\bibfield  {journal} {\bibinfo  {journal}
  {Astrophys. Space Sci.}\ }\textbf {\bibinfo {volume} {305}},\ \bibinfo
  {pages} {241} (\bibinfo {year} {2006})}\BibitemShut {NoStop}%
\bibitem [{\citenamefont {Picoli}\ \emph {et~al.}(2009)\citenamefont {Picoli},
  \citenamefont {Mendes}, \citenamefont {Malacarne},\ and\ \citenamefont
  {Santos}}]{Picoli2009}%
  \BibitemOpen
  \bibfield  {author} {\bibinfo {author} {\bibfnamefont {S.}~\bibnamefont
  {Picoli}}, \bibinfo {author} {\bibfnamefont {R.~S.}\ \bibnamefont {Mendes}},
  \bibinfo {author} {\bibfnamefont {L.~C.}\ \bibnamefont {Malacarne}}, \ and\
  \bibinfo {author} {\bibfnamefont {R.~P.~B.}\ \bibnamefont {Santos}},\
  }\href@noop {} {\bibfield  {journal} {\bibinfo  {journal} {Braz. J. Phys.}\
  }\textbf {\bibinfo {volume} {39}},\ \bibinfo {pages} {468} (\bibinfo {year}
  {2009})}\BibitemShut {NoStop}%
\bibitem [{\citenamefont {Vallianatos}(2013)}]{Vallianatos2013}%
  \BibitemOpen
  \bibfield  {author} {\bibinfo {author} {\bibfnamefont {F.}~\bibnamefont
  {Vallianatos}},\ }\href@noop {} {\bibfield  {journal} {\bibinfo  {journal}
  {EPL}\ }\textbf {\bibinfo {volume} {102}},\ \bibinfo {pages} {28006}
  (\bibinfo {year} {2013})}\BibitemShut {NoStop}%
\bibitem [{\citenamefont {Tsallis}(2009{\natexlab{b}})}]{Tsallis2009-2}%
  \BibitemOpen
  \bibfield  {author} {\bibinfo {author} {\bibfnamefont {C.}~\bibnamefont
  {Tsallis}},\ }\href@noop {} {\bibfield  {journal} {\bibinfo  {journal} {Braz.
  J. Phys.}\ }\textbf {\bibinfo {volume} {39}},\ \bibinfo {pages} {337}
  (\bibinfo {year} {2009}{\natexlab{b}})}\BibitemShut {NoStop}%
\bibitem [{\citenamefont {Hanel}\ and\ \citenamefont
  {Thurner}(2011)}]{Hanel2011-1}%
  \BibitemOpen
  \bibfield  {author} {\bibinfo {author} {\bibfnamefont {R.}~\bibnamefont
  {Hanel}}\ and\ \bibinfo {author} {\bibfnamefont {S.}~\bibnamefont
  {Thurner}},\ }\href@noop {} {\bibfield  {journal} {\bibinfo  {journal} {EPL}\
  }\textbf {\bibinfo {volume} {93}},\ \bibinfo {pages} {20006} (\bibinfo {year}
  {2011})}\BibitemShut {NoStop}%
\bibitem [{\citenamefont {Hanel}\ \emph {et~al.}(2011)\citenamefont {Hanel},
  \citenamefont {Thurner},\ and\ \citenamefont {Gell-Mann}}]{Hanel2011-2}%
  \BibitemOpen
  \bibfield  {author} {\bibinfo {author} {\bibfnamefont {R.}~\bibnamefont
  {Hanel}}, \bibinfo {author} {\bibfnamefont {S.}~\bibnamefont {Thurner}}, \
  and\ \bibinfo {author} {\bibfnamefont {M.}~\bibnamefont {Gell-Mann}},\
  }\href@noop {} {\bibfield  {journal} {\bibinfo  {journal} {PNAS}\ }\textbf
  {\bibinfo {volume} {108}},\ \bibinfo {pages} {6390} (\bibinfo {year}
  {2011})}\BibitemShut {NoStop}%
\bibitem [{\citenamefont {Ruseckas}(2016)}]{Ruseckas2016}%
  \BibitemOpen
  \bibfield  {author} {\bibinfo {author} {\bibfnamefont {J.}~\bibnamefont
  {Ruseckas}},\ }\href@noop {} {\bibfield  {journal} {\bibinfo  {journal}
  {Physica A}\ }\textbf {\bibinfo {volume} {447}},\ \bibinfo {pages} {85}
  (\bibinfo {year} {2016})}\BibitemShut {NoStop}%
\bibitem [{\citenamefont {Lutsko}\ and\ \citenamefont
  {Boon}(2011)}]{Lutsko2011}%
  \BibitemOpen
  \bibfield  {author} {\bibinfo {author} {\bibfnamefont {J.~F.}\ \bibnamefont
  {Lutsko}}\ and\ \bibinfo {author} {\bibfnamefont {J.~P.}\ \bibnamefont
  {Boon}},\ }\href@noop {} {\bibfield  {journal} {\bibinfo  {journal} {EPL}\
  }\textbf {\bibinfo {volume} {95}},\ \bibinfo {pages} {20006} (\bibinfo {year}
  {2011})}\BibitemShut {NoStop}%
\bibitem [{\citenamefont {Abe}\ and\ \citenamefont
  {Rajagopal}(2001)}]{Abe2001}%
  \BibitemOpen
  \bibfield  {author} {\bibinfo {author} {\bibfnamefont {S.}~\bibnamefont
  {Abe}}\ and\ \bibinfo {author} {\bibfnamefont {A.~K.}\ \bibnamefont
  {Rajagopal}},\ }\href@noop {} {\bibfield  {journal} {\bibinfo  {journal}
  {Europhys. Lett.}\ }\textbf {\bibinfo {volume} {55}},\ \bibinfo {pages} {6}
  (\bibinfo {year} {2001})}\BibitemShut {NoStop}%
\bibitem [{\citenamefont {Plastino}\ and\ \citenamefont
  {Plastino}(1994)}]{Plastino1994}%
  \BibitemOpen
  \bibfield  {author} {\bibinfo {author} {\bibfnamefont {A.~R.}\ \bibnamefont
  {Plastino}}\ and\ \bibinfo {author} {\bibfnamefont {A.}~\bibnamefont
  {Plastino}},\ }\href@noop {} {\bibfield  {journal} {\bibinfo  {journal}
  {Phys. Lett. A}\ }\textbf {\bibinfo {volume} {193}},\ \bibinfo {pages} {140}
  (\bibinfo {year} {1994})}\BibitemShut {NoStop}%
\bibitem [{\citenamefont {Wada}(2002)}]{Wada2002}%
  \BibitemOpen
  \bibfield  {author} {\bibinfo {author} {\bibfnamefont {T.}~\bibnamefont
  {Wada}},\ }\href@noop {} {\bibfield  {journal} {\bibinfo  {journal} {Phys.
  Lett. A}\ }\textbf {\bibinfo {volume} {297}},\ \bibinfo {pages} {334}
  (\bibinfo {year} {2002})}\BibitemShut {NoStop}%
\bibitem [{\citenamefont {Nobre}\ \emph {et~al.}(2015)\citenamefont {Nobre},
  \citenamefont {Curado}, \citenamefont {Souza},\ and\ \citenamefont
  {Andrade}}]{Nobre2015}%
  \BibitemOpen
  \bibfield  {author} {\bibinfo {author} {\bibfnamefont {F.~D.}\ \bibnamefont
  {Nobre}}, \bibinfo {author} {\bibfnamefont {E.~M.~F.}\ \bibnamefont
  {Curado}}, \bibinfo {author} {\bibfnamefont {A.~M.~C.}\ \bibnamefont
  {Souza}}, \ and\ \bibinfo {author} {\bibfnamefont {R.~F.~S.}\ \bibnamefont
  {Andrade}},\ }\href@noop {} {\bibfield  {journal} {\bibinfo  {journal} {Phys.
  Rev. E}\ }\textbf {\bibinfo {volume} {91}},\ \bibinfo {pages} {022135}
  (\bibinfo {year} {2015})}\BibitemShut {NoStop}%
\bibitem [{\citenamefont {Ou}\ and\ \citenamefont {Chen}(2006)}]{Ou2006}%
  \BibitemOpen
  \bibfield  {author} {\bibinfo {author} {\bibfnamefont {C.}~\bibnamefont
  {Ou}}\ and\ \bibinfo {author} {\bibfnamefont {J.}~\bibnamefont {Chen}},\
  }\href@noop {} {\bibfield  {journal} {\bibinfo  {journal} {Physica A}\
  }\textbf {\bibinfo {volume} {370}},\ \bibinfo {pages} {525} (\bibinfo {year}
  {2006})}\BibitemShut {NoStop}%
\bibitem [{\citenamefont {Scarfone}(2010)}]{Scarfone2010}%
  \BibitemOpen
  \bibfield  {author} {\bibinfo {author} {\bibfnamefont {A.~M.}\ \bibnamefont
  {Scarfone}},\ }\href@noop {} {\bibfield  {journal} {\bibinfo  {journal}
  {Phys. Lett. A}\ }\textbf {\bibinfo {volume} {374}},\ \bibinfo {pages} {2701}
  (\bibinfo {year} {2010})}\BibitemShut {NoStop}%
\bibitem [{\citenamefont {Thirring}(1983)}]{Thirring1983}%
  \BibitemOpen
  \bibfield  {author} {\bibinfo {author} {\bibfnamefont {W.}~\bibnamefont
  {Thirring}},\ }\href@noop {} {\emph {\bibinfo {title} {Quantum Mechanics of
  Large Systems}}}\ (\bibinfo  {publisher} {Springer-Verlag},\ \bibinfo
  {address} {New York},\ \bibinfo {year} {1983})\BibitemShut {NoStop}%
\bibitem [{\citenamefont {Posch}\ \emph {et~al.}(1990)\citenamefont {Posch},
  \citenamefont {Narnhofer},\ and\ \citenamefont {Thirring}}]{Posch1990}%
  \BibitemOpen
  \bibfield  {author} {\bibinfo {author} {\bibfnamefont {H.~A.}\ \bibnamefont
  {Posch}}, \bibinfo {author} {\bibfnamefont {H.}~\bibnamefont {Narnhofer}}, \
  and\ \bibinfo {author} {\bibfnamefont {W.}~\bibnamefont {Thirring}},\
  }\href@noop {} {\bibfield  {journal} {\bibinfo  {journal} {Phys. Rev. A}\
  }\textbf {\bibinfo {volume} {42}},\ \bibinfo {pages} {1880} (\bibinfo {year}
  {1990})}\BibitemShut {NoStop}%
\bibitem [{\citenamefont {Posch}\ \emph {et~al.}(1993)\citenamefont {Posch},
  \citenamefont {Narnhofer},\ and\ \citenamefont {Thirring}}]{Posch1993}%
  \BibitemOpen
  \bibfield  {author} {\bibinfo {author} {\bibfnamefont {H.~A.}\ \bibnamefont
  {Posch}}, \bibinfo {author} {\bibfnamefont {H.}~\bibnamefont {Narnhofer}}, \
  and\ \bibinfo {author} {\bibfnamefont {W.}~\bibnamefont {Thirring}},\
  }\href@noop {} {\bibfield  {journal} {\bibinfo  {journal} {Physica A}\
  }\textbf {\bibinfo {volume} {194}},\ \bibinfo {pages} {482} (\bibinfo {year}
  {1993})}\BibitemShut {NoStop}%
\bibitem [{\citenamefont {Latora}\ \emph {et~al.}(1998)\citenamefont {Latora},
  \citenamefont {Rapisarda},\ and\ \citenamefont {Ruffo}}]{Latora1998}%
  \BibitemOpen
  \bibfield  {author} {\bibinfo {author} {\bibfnamefont {V.}~\bibnamefont
  {Latora}}, \bibinfo {author} {\bibfnamefont {A.}~\bibnamefont {Rapisarda}}, \
  and\ \bibinfo {author} {\bibfnamefont {S.}~\bibnamefont {Ruffo}},\
  }\href@noop {} {\bibfield  {journal} {\bibinfo  {journal} {Phys. Rev. Lett.}\
  }\textbf {\bibinfo {volume} {80}},\ \bibinfo {pages} {692} (\bibinfo {year}
  {1998})}\BibitemShut {NoStop}%
\bibitem [{\citenamefont {Antoni}\ and\ \citenamefont
  {Torcini}(1998)}]{Antoni1998}%
  \BibitemOpen
  \bibfield  {author} {\bibinfo {author} {\bibfnamefont {M.}~\bibnamefont
  {Antoni}}\ and\ \bibinfo {author} {\bibfnamefont {A.}~\bibnamefont
  {Torcini}},\ }\href@noop {} {\bibfield  {journal} {\bibinfo  {journal} {Phys.
  Rev. E}\ }\textbf {\bibinfo {volume} {57}},\ \bibinfo {pages} {R6233}
  (\bibinfo {year} {1998})}\BibitemShut {NoStop}%
\bibitem [{\citenamefont {Dauxois}\ \emph {et~al.}(2002)\citenamefont
  {Dauxois}, \citenamefont {Ruffo}, \citenamefont {Arimondo},\ and\
  \citenamefont {Wilkens}}]{Dauxois2002}%
  \BibitemOpen
  \bibinfo {editor} {\bibfnamefont {T.}~\bibnamefont {Dauxois}}, \bibinfo
  {editor} {\bibfnamefont {S.}~\bibnamefont {Ruffo}}, \bibinfo {editor}
  {\bibfnamefont {E.}~\bibnamefont {Arimondo}}, \ and\ \bibinfo {editor}
  {\bibfnamefont {M.}~\bibnamefont {Wilkens}},\ eds.,\ \href@noop {} {\emph
  {\bibinfo {title} {Dynamics and Thermodynamics of Systems with Long-Range
  Interactions}}},\ \bibinfo {series} {Lect. Not. Phys.}, Vol.\ \bibinfo
  {volume} {602}\ (\bibinfo  {publisher} {Springer},\ \bibinfo {address} {New
  York},\ \bibinfo {year} {2002})\BibitemShut {NoStop}%
\bibitem [{\citenamefont {Barr{\'e}}\ \emph {et~al.}(2001)\citenamefont
  {Barr{\'e}}, \citenamefont {Mukamel},\ and\ \citenamefont
  {Ruffo}}]{Barre2001}%
  \BibitemOpen
  \bibfield  {author} {\bibinfo {author} {\bibfnamefont {J.}~\bibnamefont
  {Barr{\'e}}}, \bibinfo {author} {\bibfnamefont {D.}~\bibnamefont {Mukamel}},
  \ and\ \bibinfo {author} {\bibfnamefont {S.}~\bibnamefont {Ruffo}},\
  }\href@noop {} {\bibfield  {journal} {\bibinfo  {journal} {Phys. Rev. Lett.}\
  }\textbf {\bibinfo {volume} {87}},\ \bibinfo {pages} {030601} (\bibinfo
  {year} {2001})}\BibitemShut {NoStop}%
\bibitem [{\citenamefont {Mukamel}\ \emph {et~al.}(2005)\citenamefont
  {Mukamel}, \citenamefont {Ruffo},\ and\ \citenamefont
  {Schreiber}}]{Mukamel2005}%
  \BibitemOpen
  \bibfield  {author} {\bibinfo {author} {\bibfnamefont {D.}~\bibnamefont
  {Mukamel}}, \bibinfo {author} {\bibfnamefont {S.}~\bibnamefont {Ruffo}}, \
  and\ \bibinfo {author} {\bibfnamefont {N.}~\bibnamefont {Schreiber}},\
  }\href@noop {} {\bibfield  {journal} {\bibinfo  {journal} {Phys. Rev. Lett.}\
  }\textbf {\bibinfo {volume} {95}},\ \bibinfo {pages} {240604} (\bibinfo
  {year} {2005})}\BibitemShut {NoStop}%
\bibitem [{\citenamefont {Lynden-Bell}\ and\ \citenamefont
  {Wood}(1968)}]{Lynden-Bell1968}%
  \BibitemOpen
  \bibfield  {author} {\bibinfo {author} {\bibfnamefont {D.}~\bibnamefont
  {Lynden-Bell}}\ and\ \bibinfo {author} {\bibfnamefont {R.}~\bibnamefont
  {Wood}},\ }\href@noop {} {\bibfield  {journal} {\bibinfo  {journal} {Mon.
  Not. R. Astron. Soc.}\ }\textbf {\bibinfo {volume} {138}},\ \bibinfo {pages}
  {495} (\bibinfo {year} {1968})}\BibitemShut {NoStop}%
\bibitem [{\citenamefont {Thirring}(1970)}]{Thirring1970}%
  \BibitemOpen
  \bibfield  {author} {\bibinfo {author} {\bibfnamefont {W.}~\bibnamefont
  {Thirring}},\ }\href@noop {} {\bibfield  {journal} {\bibinfo  {journal} {Z.
  Phys.}\ }\textbf {\bibinfo {volume} {235}},\ \bibinfo {pages} {339} (\bibinfo
  {year} {1970})}\BibitemShut {NoStop}%
\bibitem [{\citenamefont {Saslaw}(1985)}]{Saslaw1985}%
  \BibitemOpen
  \bibfield  {author} {\bibinfo {author} {\bibfnamefont {W.~C.}\ \bibnamefont
  {Saslaw}},\ }\href@noop {} {\emph {\bibinfo {title} {Gravitational Physics of
  Stellar and Galactic Systems}}}\ (\bibinfo  {publisher} {Cambridge University
  Press},\ \bibinfo {address} {Cambridge},\ \bibinfo {year} {1985})\BibitemShut
  {NoStop}%
\bibitem [{\citenamefont {Bixon}\ and\ \citenamefont
  {Jortner}(1989)}]{Bixon1989}%
  \BibitemOpen
  \bibfield  {author} {\bibinfo {author} {\bibfnamefont {M.}~\bibnamefont
  {Bixon}}\ and\ \bibinfo {author} {\bibfnamefont {J.}~\bibnamefont
  {Jortner}},\ }\href@noop {} {\bibfield  {journal} {\bibinfo  {journal} {J.
  Chem. Phys.}\ }\textbf {\bibinfo {volume} {91}},\ \bibinfo {pages} {1631}
  (\bibinfo {year} {1989})}\BibitemShut {NoStop}%
\bibitem [{\citenamefont {Labastie}\ and\ \citenamefont
  {Whetten}(1990)}]{Labastie1990}%
  \BibitemOpen
  \bibfield  {author} {\bibinfo {author} {\bibfnamefont {P.}~\bibnamefont
  {Labastie}}\ and\ \bibinfo {author} {\bibfnamefont {R.~L.}\ \bibnamefont
  {Whetten}},\ }\href@noop {} {\bibfield  {journal} {\bibinfo  {journal} {Phys.
  Rev. Lett.}\ }\textbf {\bibinfo {volume} {65}},\ \bibinfo {pages} {1567}
  (\bibinfo {year} {1990})}\BibitemShut {NoStop}%
\bibitem [{\citenamefont {Gross}(1990)}]{Gross1990}%
  \BibitemOpen
  \bibfield  {author} {\bibinfo {author} {\bibfnamefont {D.~H.~E.}\
  \bibnamefont {Gross}},\ }\href@noop {} {\bibfield  {journal} {\bibinfo
  {journal} {Rep. Prog. Phys.}\ }\textbf {\bibinfo {volume} {53}},\ \bibinfo
  {pages} {605} (\bibinfo {year} {1990})}\BibitemShut {NoStop}%
\bibitem [{\citenamefont {D'Agostino}\ \emph {et~al.}(2000)\citenamefont
  {D'Agostino}, \citenamefont {Gulminelli}, \citenamefont {Chomaz},
  \citenamefont {Bruno}, \citenamefont {Cannata}, \citenamefont {Bougault},
  \citenamefont {Gramegna}, \citenamefont {Iori}, \citenamefont {Neindre},
  \citenamefont {Margagliotti}, \citenamefont {Moroni},\ and\ \citenamefont
  {Vannini}}]{DAgostino2000}%
  \BibitemOpen
  \bibfield  {author} {\bibinfo {author} {\bibfnamefont {M.}~\bibnamefont
  {D'Agostino}}, \bibinfo {author} {\bibfnamefont {F.}~\bibnamefont
  {Gulminelli}}, \bibinfo {author} {\bibfnamefont {P.}~\bibnamefont {Chomaz}},
  \bibinfo {author} {\bibfnamefont {M.}~\bibnamefont {Bruno}}, \bibinfo
  {author} {\bibfnamefont {F.}~\bibnamefont {Cannata}}, \bibinfo {author}
  {\bibfnamefont {R.}~\bibnamefont {Bougault}}, \bibinfo {author}
  {\bibfnamefont {F.}~\bibnamefont {Gramegna}}, \bibinfo {author}
  {\bibfnamefont {I.}~\bibnamefont {Iori}}, \bibinfo {author} {\bibfnamefont
  {N.~L.}\ \bibnamefont {Neindre}}, \bibinfo {author} {\bibfnamefont {G.~V.}\
  \bibnamefont {Margagliotti}}, \bibinfo {author} {\bibfnamefont
  {A.}~\bibnamefont {Moroni}}, \ and\ \bibinfo {author} {\bibfnamefont
  {G.}~\bibnamefont {Vannini}},\ }\href@noop {} {\bibfield  {journal} {\bibinfo
   {journal} {Phys. Lett. B}\ }\textbf {\bibinfo {volume} {473}},\ \bibinfo
  {pages} {219} (\bibinfo {year} {2000})}\BibitemShut {NoStop}%
\bibitem [{\citenamefont {Schmidt}\ \emph {et~al.}(2001)\citenamefont
  {Schmidt}, \citenamefont {Kusche}, \citenamefont {Hippler}, \citenamefont
  {Donges}, \citenamefont {Kronm{\"u}ller}, \citenamefont {von Issendorff},\
  and\ \citenamefont {Haberland}}]{Schmidt2001}%
  \BibitemOpen
  \bibfield  {author} {\bibinfo {author} {\bibfnamefont {M.}~\bibnamefont
  {Schmidt}}, \bibinfo {author} {\bibfnamefont {R.}~\bibnamefont {Kusche}},
  \bibinfo {author} {\bibfnamefont {T.}~\bibnamefont {Hippler}}, \bibinfo
  {author} {\bibfnamefont {J.}~\bibnamefont {Donges}}, \bibinfo {author}
  {\bibfnamefont {W.}~\bibnamefont {Kronm{\"u}ller}}, \bibinfo {author}
  {\bibfnamefont {B.}~\bibnamefont {von Issendorff}}, \ and\ \bibinfo {author}
  {\bibfnamefont {H.}~\bibnamefont {Haberland}},\ }\href@noop {} {\bibfield
  {journal} {\bibinfo  {journal} {Phys. Rev. Lett.}\ }\textbf {\bibinfo
  {volume} {86}},\ \bibinfo {pages} {1191} (\bibinfo {year}
  {2001})}\BibitemShut {NoStop}%
\bibitem [{\citenamefont {Abe}(1999)}]{Abe1999a}%
  \BibitemOpen
  \bibfield  {author} {\bibinfo {author} {\bibfnamefont {S.}~\bibnamefont
  {Abe}},\ }\href@noop {} {\bibfield  {journal} {\bibinfo  {journal} {Phys.
  Lett. A}\ }\textbf {\bibinfo {volume} {263}},\ \bibinfo {pages} {424}
  (\bibinfo {year} {1999})}\BibitemShut {NoStop}%
\bibitem [{\citenamefont {Abe}(2000)}]{Abe2000}%
  \BibitemOpen
  \bibfield  {author} {\bibinfo {author} {\bibfnamefont {S.}~\bibnamefont
  {Abe}},\ }\href@noop {} {\bibfield  {journal} {\bibinfo  {journal} {Phys.
  Lett. A}\ }\textbf {\bibinfo {volume} {267}},\ \bibinfo {pages} {456}
  (\bibinfo {year} {2000})}\BibitemShut {NoStop}%
\bibitem [{\citenamefont {{Di Sisto}}\ \emph {et~al.}(1999)\citenamefont {{Di
  Sisto}}, \citenamefont {Mart{\'\i}nez}, \citenamefont {Orellana},
  \citenamefont {Plastino},\ and\ \citenamefont {Plastino}}]{DiSisto1999}%
  \BibitemOpen
  \bibfield  {author} {\bibinfo {author} {\bibfnamefont {R.~P.}\ \bibnamefont
  {{Di Sisto}}}, \bibinfo {author} {\bibfnamefont {S.}~\bibnamefont
  {Mart{\'\i}nez}}, \bibinfo {author} {\bibfnamefont {R.~B.}\ \bibnamefont
  {Orellana}}, \bibinfo {author} {\bibfnamefont {A.~R.}\ \bibnamefont
  {Plastino}}, \ and\ \bibinfo {author} {\bibfnamefont {A.}~\bibnamefont
  {Plastino}},\ }\href@noop {} {\bibfield  {journal} {\bibinfo  {journal}
  {Physica A}\ }\textbf {\bibinfo {volume} {265}},\ \bibinfo {pages} {590}
  (\bibinfo {year} {1999})}\BibitemShut {NoStop}%
\bibitem [{\citenamefont {Plastino}\ and\ \citenamefont
  {Rocca}(2013)}]{Plastino2013}%
  \BibitemOpen
  \bibfield  {author} {\bibinfo {author} {\bibfnamefont {A.}~\bibnamefont
  {Plastino}}\ and\ \bibinfo {author} {\bibfnamefont {M.~C.}\ \bibnamefont
  {Rocca}},\ }\href@noop {} {\bibfield  {journal} {\bibinfo  {journal} {EPL}\
  }\textbf {\bibinfo {volume} {104}},\ \bibinfo {pages} {60003} (\bibinfo
  {year} {2013})}\BibitemShut {NoStop}%
\bibitem [{\citenamefont {Lutsko}\ and\ \citenamefont
  {Boon}(2014)}]{Lutsko2014}%
  \BibitemOpen
  \bibfield  {author} {\bibinfo {author} {\bibfnamefont {J.~F.}\ \bibnamefont
  {Lutsko}}\ and\ \bibinfo {author} {\bibfnamefont {J.~P.}\ \bibnamefont
  {Boon}},\ }\href@noop {} {\bibfield  {journal} {\bibinfo  {journal} {EPL}\
  }\textbf {\bibinfo {volume} {107}},\ \bibinfo {pages} {10003} (\bibinfo
  {year} {2014})}\BibitemShut {NoStop}%
\end{thebibliography}
\end{document}